\documentclass[aps,prl,twocolumn,showpacs,amsmath,amssymb]{revtex4}

\usepackage{graphicx}% Include figure files
\usepackage{dcolumn}% Align table columns on decimal point
\usepackage{bm}% bold math

\renewcommand{\vec}[1]{\mathbf{#1}}

\newcommand{\degree}{^{\circ}}

\begin{document}

\def\bs{\boldsymbol}

\title{Faraday Wave Pattern Selection Via Multi-Frequency Forcing}
\author{Jeff Porter}
\email{jport@maths.leeds.ac.uk}
\affiliation{Department of Applied Mathematics, University of Leeds, 
Leeds LS2 9JT, United Kingdom}
\author{Chad M. Topaz}
\affiliation{Department of Mathematics, U.C.L.A., Los Angeles, California 90095}
\author{Mary Silber}
\affiliation{Department of Engineering Sciences and Applied Mathematics,
Northwestern University, Evanston, Illinois 60208}

\date{\today}

\begin{abstract}
We use symmetry considerations to investigate how damped modes affect pattern 
selection in multi-frequency forced Faraday waves. We classify and tabulate the 
most important damped modes and determine how the corresponding resonant 
triad interactions depend on the forcing parameters. The relative phase of the 
forcing terms may be used to enhance or suppress the nonlinear interactions. We 
compare our predictions with numerical results and discuss their implications for 
recent experiments. Our results suggest how to design multi-frequency forcing 
functions that favor chosen patterns in the lab.
\end{abstract}


\maketitle

Faraday observed  in 1831 that patterns of subharmonic standing waves form on the 
surface of a fluid when its supporting container is shaken vertically with increasing 
strength \cite{Far1831}.  The first Faraday wave experiments (see 
\cite{MilHen90} for a review) used sinusoidal forcing to shake the fluid, 
and produced simple patterns (\emph{e.g.}, stripes, squares, and hexagons). Remarkably, 
this experiment remains a rich source of intriguing new patterns, most recently as the 
basic sinusoidal forcing has been replaced by periodic functions with two or three 
frequency components.   The two-frequency case has been investigated a great deal 
experimentally \cite{Mul93,EdwFau94,ArbFin98,KudPieGol98,ArbFin02} and 
theoretically \cite{BesEdwTuc96,ZhaVin97_2,SilTopSke00,PorSil02}, and has produced 
a variety of states, including superlattice patterns and localized structures, as well as the 
first experimental realizations of quasipatterns.  Three-frequency forcing has been used 
too \cite{Mul93,ArbFin02}, but to far less extent.

\begin{figure}
\includegraphics[width=2.6in]{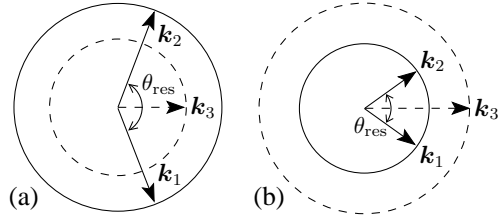}
\caption{Fourier space diagram of resonant triads composed of two critical modes 
($|\vec{k}_1|=|\vec{k}_2|=k_1$) and a damped mode ($\vec{k}_3=\vec{k}_1+\vec{k}_2$, 
$|\vec{k}_3|=k_3$) with (a) $k_3 <k_1$, (b) $k_1 < k_3 < 2k_1$. The resonant angle is 
given by $\cos (\theta_{\rm res}/2) = k_3/(2k_1)$.}
\label{fig:restriad}
\end{figure}

A primary focus of the past 10 years of investigation of one- and two-frequency forced 
Faraday waves has been on the role of resonant triad interactions -- the lowest order 
nonlinear interactions -- in pattern selection.  A typical resonant triad consists of two 
critical modes and a third, damped mode with a distinct wavelength that determines 
the angle $\theta_{\rm res}$ between the critical wave vectors; see Fig.~\ref{fig:restriad}. 
It was originally thought that the damped mode should draw energy from the excited modes, 
creating an ``anti-selection'' mechanism that suppresses such triads in favor of patterns that 
avoid the resonant angle \cite{ZhaVin97_1}. However, it has since been shown that the 
opposite effect may also occur. For instance, \cite{SilTopSke00,TopSil02} demonstrated 
for two-frequency forced Faraday waves that a damped mode oscillating at the difference 
of the two forcing frequencies is responsible for selecting the superlattice-I pattern 
reported in~\cite{KudPieGol98}.

%%%%%%%%%%%%%%%%%%%%%%%%%%%%%%%%%%%%%%%%%%%%
%%%%%%%%%%%%%%%%%%%%%%%%%%%%%%%%%%%%%%%%%%%%

The richness of the Faraday system is due in large part to the vastness of the control 
parameter space which, in principle, is infinite-dimensional. By expanding an arbitrary 
periodic forcing function in a Fourier series
\begin{align}
F(t) &= f_m e^{im\omega t} + f_n e^{i n \omega t} + f_p e^{i p \omega t} + \cdots + c.c.,
\end{align}
it is apparent that one has the freedom to choose the commensurate frequencies 
$(m\omega,n\omega,p\omega,\ldots)$, the amplitudes $(|f_m|,|f_n|,|f_p|,\ldots)$, and the phases 
$(\phi_m,\phi_n,\phi_p,\ldots)$, where $\phi_u = \arg(f_u)$. Previous experimental and 
theoretical investigations have largely been exploratory in nature, \emph{i.e.}, they have 
focused on describing the patterns which form at particular locations in parameter 
space. In this Letter, we take a prescriptive approach to pattern formation and 
demonstrate how the huge parameter space may be exploited to control resonant triads.  
In particular, we can either enhance or suppress triad interactions, largely at 
our discretion, by an appropriate choice of forcing frequencies, amplitudes, and phases.

Our analysis follows that of \cite{PorSil02} which uses the (broken) symmetries of time 
translation, time reversal, and Hamiltonian structure, and is valid near onset in weakly-damped 
systems. Our main result is a table summarizing, for up to three forcing frequencies, which 
damped modes are likely to be important, the manner in which their coupling with critical 
modes depends on the forcing parameters and, in many cases, the overall qualitative effect 
(enhancing or suppressing) they have on associated patterns. These results are then applied 
to two specific examples relevant to experiments. In the first, we show how the superlattice 
pattern selection mechanism discussed in \cite{SilTopSke00,TopSil02} may be dramatically 
enhanced by the addition of a third frequency component with appropriate phase. In the 
second, we offer an explanation for why a distorted 8-fold quasipattern observed with two-frequency 
forcing became much more regular and robust when a third forcing frequency component was 
applied in the experiments of \cite{ArbFin02}.

We consider resonant triads composed of two critical modes  and a third linearly damped 
mode (which may, nevertheless, be forced) oscillating with dominant frequency $\Omega > 0$.  
The particular values of $\Omega$ most crucial to pattern selection will be determined in the 
course of our calculation. Time is rescaled so that the common frequency is one, and we take 
the $m$ component of the forcing to drive the critical modes, which therefore have dominant 
frequency $m/2$. All other modes are linearly damped.  We begin by expanding the fluid surface 
height $h$  in terms of six traveling wave (TW) modes:
\begin{align}
h({\bs x},t) = \sum_{j=1}^3\sum_{\pm} Z_j^\pm (t)\,e^{i({\bs k_j}\cdot 
{\bs x}\pm \varpi_j t)} + c.c. + ...\,,
\label{eq:h}
\end{align} 
where $\varpi_1=\varpi_2=m/2$ and $\varpi_3=\Omega$.  The TW amplitude equations respect 
the spatial symmetries of the problem (translation, reflection through ${\bs k}_3$, inversion through 
the origin) and the temporal symmetries
\begin{subequations}
\label{eq:tempsyms} 
\begin{align}   
T_\tau &: Z_j^{\pm} \rightarrow e^{\pm i \varpi_j \tau}Z_j^{\pm},
\:\: f_u \rightarrow e^{i u \tau} f_u , \\ 
\kappa &: Z_j^{\pm} \leftrightarrow Z_j^{\mp},\:\: (t,\gamma) \rightarrow -(t,\gamma),
\:\: f_u \rightarrow \bar{f}_u\,.
\end{align}
\end{subequations} 
Here $u$ denotes any of the frequencies $\{m,n,p,\cdots\}$ and  $\gamma = 2 \nu k^2$ 
is a dimensionless damping parameter formed from the kinematic viscosity $\nu$ and the 
characteristic wave number $k(m/2)$, determined by the hydrodynamic dispersion 
relation.   Note that $T_\tau$ and $\kappa$ are not symmetries in the usual sense; time 
translation $T_{\tau}$ is broken by the forcing and time reversal $\kappa$ by the damping.  
Nonetheless, they are valid {\it parameter symmetries} of the problem.

The equivariant TW equations take the form
\begin{subequations}
\label{eq:TW}
\begin{align}
\dot{Z}_1^+ & =  \upsilon Z_1^+  - \lambda f_m Z_1^-  
+ {\cal Q}_1(Z_2^\pm,Z_3^\pm) + \cdots,\\
\dot{Z}_3^+ & =  \varrho Z_1^+  - \mu F_{2\Omega} Z_3^-  
+ {\cal Q}_3(Z_1^\pm,Z_2^\pm) + \cdots,
\end{align}
\end{subequations}
with the remaining four equations related by symmetry.  The parametric forcing term 
$F_{2\Omega}$ represents products of the $f_u$ whose frequencies sum to $2\Omega$ 
($F_{2\Omega}=f_{2\Omega}$ when $2\Omega$ forcing is present).  The resonant terms 
\begin{align*}
{\cal Q}_1 & = (Q_1Z_3^+ + Q_2Z_3^-)\bar{Z}_2^+ + (Q_3Z_3^+ + Q_4Z_3^-)\bar{Z}_2^- ,\\
{\cal Q}_3 & = Q_5 Z_1^+Z_2^+ + Q_6(Z_1^+Z_2^- + Z_1^-Z_2^+) + Q_7Z_1^-Z_2^-,   
\end{align*}
couple the two modes at quadratic order.  The resonant coefficients $Q_\ell$, according to $T_\tau$, must 
transform as $(Q_1,\bar{Q}_5) \sim e^{i(m-\Omega)\tau}$,  $(Q_2,Q_7) \sim e^{i(m+\Omega)\tau}$, 
$(Q_3,\bar{Q}_4,\bar{Q}_6) \sim e^{-i\Omega \tau}$.  The dependence of the $T_\tau$-invariant 
coefficients $\upsilon$, $\lambda$, etc., on $\gamma$ and the $f_u$, is also determined by 
symmetry.   In particular, $\kappa$ forces most of them to be purely imaginary at
leading order.  However, if $k_1$ and $k_3$ are defined by the local minima of the neutral stability 
curves then the  imaginary parts of $\upsilon$ and $\varrho$ (\emph{i.e.}, the detunings) vanish 
at $\gamma =0$.  We will need only the leading term in each coefficient so we write
\begin{align*}
&\upsilon = -\upsilon_r \gamma, \;\; \varrho = -\varrho_r \gamma,  \;\;
\lambda = i \lambda_i,  \;\; \mu= i \mu_i,  \;\; Q_\ell = i q_\ell F_\ell ,
\end{align*}
where $\upsilon_r, \varrho_r> 0$ (they correspond to damping terms), $\lambda_i,\mu_i>0$ 
(see \cite{PorSil_pre}), $q_\ell \in \mathbb{R}$ and $F_\ell$ denotes an appropriate product 
of the $f_u$ (or unity).    The critical value of $|f_m|$ is ${\cal O}(\gamma)$ 
and all $|f_u|$ are assumed to be related by a finite $\gamma$-independent ratio, so that 
$|f_u| \sim \gamma$.  

For a given damped mode, the task is to determine how the resonant coefficients $Q_\ell$ 
depend on the $f_u$, then apply a standard reduction procedure at the bifurcation to standing 
waves (SW).   The resulting two equations (truncated at cubic order) describe SW dynamics 
near onset: 
\begin{align}
\dot{A}_1 = \lambda A_1 + A_1\! \left(a|A_1|^2 + (b_0 + b_{\rm res})|A_2|^2\right). 
\label{eq:SW}
\end{align}
The equation for $A_2$ is obtained by switching labels.   The self-interaction coefficient $a$ 
and the ``nonresonant'' part $b_0$ of the cross-coupling coefficient are ${\cal O}(\gamma)$.

Of primary interest here is the contribution $b_{\rm res}$ makes to the cross-coupling 
coefficient in (\ref{eq:SW}) as a result of the slaved modes $Z_3^\pm$.  Loosely 
speaking, if $b_{\rm res}>0$ the stability of patterns involving critical modes separated by 
the angle $\theta_{\rm res}$ will be enhanced, whereas $b_{\rm res}<0$ has a suppressing 
effect \cite{SilTopSke00}.  Since we are interested only in damped modes that have a 
significant impact on (\ref{eq:SW}), we consider those cases where $b_{\rm res}$ is 
${\cal O}(\gamma)$ or larger \footnote{One can amplify the effect of modes with weaker 
coupling by bringing them extremely close to onset but this is effectively a {\it bicritical} 
situation which we avoid here.}. 

There are then two possibilities for obtaining large $b_{\rm res}$. If some of the $Q_\ell$ 
are ${\cal O}(1)$ then $b_{\rm res}$ is ${\cal O}(\gamma^{-1})$ and dominates in the limit 
$\gamma \rightarrow 0$; this happens only for $\Omega = m$ and is in essence a single-frequency 
phenomenon (the ``first harmonic" resonance).   If $\Omega \neq m$ and some of the $Q_\ell$ 
are ${\cal O}(\gamma)$ then $b_{\rm res}$ is ${\cal O}(\gamma)$, and hence comparable to 
$b_0$.  This occurs when $\Omega \in \{2m, n, m \pm n, n-m \}$ for some frequency $n$.   
However, the $\Omega=2m$ mode may be ignored because for Faraday waves it has a wavenumber 
$k_3 > 2k_1$ and so it cannot satisfy the necessary {\it spatial} resonance condition. Each of the 
remaining conditions on $\Omega$ generates a particular type of coupling.  For example, with 
$\Omega = n-m$ we have  $F_2 = F_7 = f_n$.  If more than one condition is satisfied, as when 
$\Omega = n-m = m-p$, there are more coupling terms (here one would also have 
$F_ 1 = \bar{F}_5 = f_p$).  The maximum number of conditions that $\Omega$ can satisfy 
simultaneously is three.

In addition to the issue of coupling terms, it matters whether  or not the damped mode is parametrically 
forced at ${\cal O}(\gamma)$.  This forcing is present when there is a frequency $2\Omega$ in $F(t)$ 
and magnifies the resonance effect as it brings the damped mode closer to criticality.   In addition, it 
leads to interesting phase dependence as the ``usual" phase, dictated by $f_{2\Omega}$, competes with 
the phase of the nonlinear forcing ${\cal Q}_3$ (see Eqs.~\ref{eq:TW}).

In Table \ref{tab:3ff} we give the leading contribution to $b_{\rm res}$ of the important damped modes 
for different choices of $F(t)$ containing up to three frequencies $(m,n,p)$.
%%%%%%%%%%%%%%%%%%%%%%%%%%%%%%%%%%%%%%%%%
%%%%%%%%%%%%%%%%%%%%%%%%%%%%%%%%%%%%%%%%%
\begingroup
\squeezetable
\begin{table*}
\caption{\label{tab:3ff}Leading resonant contribution $b_{\rm res}$ in (\ref{eq:SW}) for the 
most important damped modes under appropriate choice of three-frequency forcing. Here, 
$m,n,p,\Omega > 0$ and $x\in\mathbb{Z}^+$.  Each expression for $(m,n,p)$, given $\Omega$, is 
excluded from those of entries further down the table. Dots in the first column indicate an arbitrary 
commensurate frequency. For $\star$ the $\pm$ follows sign($m-n$).}
%\begin{footnotesize}
\begin{tabular*}{\textwidth}{@{\extracolsep{\fill}}lccl}
\hline \hline
$(m,n,p)$ & $\Omega$ & Leading resonant contribution $b_{\rm res}$ & relevant phase(s) \\ 
\hline \hline
$(m,\cdot, \cdot)$ &$m$ &$-\alpha_1/|\varrho|$ & \\
$(m,n, \cdot)$ &$n$ &$-\alpha_3|f_n|^2/|\varrho|$ & \\ 
$(m,n, \cdot)$ &$m \pm n$ &$-\alpha_1|f_n|^2/|\varrho|$ & \\ 
$(m,n, \cdot)$ &$n-m$ &$\alpha_2|f_n|^2/|\varrho|$ & \\ 
\hline \hline
$(m,2m, \cdot)$ &$m$ &$-\alpha_1P_n(\Phi)$ &$\Phi=\phi_n-2\phi_m$\\ 
$(m,n,2n)$ &$n$ &$-\alpha_3|f_n|^2P_p(\Phi)$ &$\Phi=2\phi_n-\phi_p$ \\ 
$(3x,2x, \cdot)$ &$x$ &$-\alpha_1|f_n|^2P_n(\Phi)$ &$\Phi=3\phi_n-2\phi_m$ \\ 
$(m,n,2m \pm 2n)$ &$m \pm n$ &$-\alpha_1|f_n|^2P_p(\Phi)$ &$\Phi=\phi_p -2\phi_m \mp 2\phi_n$ \\ 
$(m,n,2n-2m)$ &$n-m$ &$\alpha_2|f_n|^2P_p(\Phi)$ &$\Phi=\phi_p + 2\phi_m - 2\phi_n $\\ 
\hline \hline
$(m,n,|m-n|)$ &$n$ &$(-\alpha_1|f_p|^2- \alpha_3|f_n|^2+ \alpha_5|f_n||f_p|\sin\Phi)/|\varrho|$ 
&$\Phi=\phi_n - \phi_m \pm \phi_p \,\star$ \\ 
$(m,n,m+n)$ &$n$ &$(\alpha_2|f_p|^2- \alpha_3|f_n|^2 + \alpha_6|f_n||f_p|\sin\Phi)/|\varrho|$ 
&$\Phi=\phi_m + \phi_n - \phi_p$ \\ 
$(m,n,2m \pm n)$ &$m \pm n$ &$(\alpha_2|f_p|^2 -\alpha_1|f_n|^2 + \alpha_4|f_n||f_p|\cos\Phi)/|\varrho|$ 
&$\Phi=2 \phi_m - \phi_p \pm \phi_n$ \\ 
\hline \hline
$(3,1,2)$ &$1$ &$-\alpha_1|f_p|^2 P_p(\Phi_1 - \Phi_2) - \alpha_3|f_n|^2 P_p(\Phi_1 + \Phi_2)$ 
&$\Phi_1=\phi_n  - \phi_m  + \phi_p$ \\
& &$\mbox{}+ \alpha_5|f_n||f_p|R_p(\Phi_1,\Phi_2)$ &$\Phi_2=\phi_m + \phi_n - 2\phi_p$ \\ 
$(3,2,4)$ &$1$&$-\alpha_1|f_n|^2 P_n(\Phi_1 + \Phi_2) + \alpha_2|f_p|^2 P_n(\Phi_2 - \Phi_1)$ 
&$\Phi_1=\phi_n + \phi_p - 2\phi_m$ \\
& &$+ \alpha_4|f_n||f_p|R_n(\Phi_1 - 90\degree,\Phi_2 + 90\degree)$ &$\Phi_2=2\phi_n - \phi_p$ \\ 
\hline \hline
\end{tabular*}
%\end{footnotesize}
\end{table*}
\endgroup
%%%%%%%%%%%%%%%%%%%%%%%%%%%%%%%%%%%%%%%%%
%%%%%%%%%%%%%%%%%%%%%%%%%%%%%%%%%%%%%%%%%
To simplify the expressions therein, we define
\begin{align}
\label{eq:alphas}
&\alpha_1 = q_{1}q_{5}, \quad \alpha_2 = q_{2}q_{7}, \quad \alpha_3 
= 2q_{6}(q_{3}-q_{4}), \nonumber \\ 
&\alpha_4 = q_{1}q_{7}-q_{2}q_{5}, \quad 
\alpha_5 = \epsilon_\lambda (2q_{1}q_{6}+q_{5}(q_{3}-q_{4})), \nonumber \\
&\alpha_6 = \epsilon_\lambda ( 2q_{2}q_{6}-q_{7}(q_{3}-q_{4})), 
\quad \epsilon_\lambda = {\rm sign}(\lambda_i),
\end{align}
and the functions
\begin{align}
&P_{2\Omega}(\Phi)= \frac{(|\varrho|+\mu_i|f_{2\Omega}|\sin\Phi)}
{(|\varrho|^2-|\mu_if_{2\Omega}|^2)}, \label{eq:P}\\
&R_{2\Omega}(\Phi_1,\Phi_2)= \frac{(|\varrho|\sin\Phi_1+\mu_i|f_{2\Omega}|\cos\Phi_2)}
{(|\varrho|^2-|\mu_i f_{2\Omega}|^2)}. \label{eq:R}
\end{align}

There are four groupings in the table. The first shows the five important damped modes 
and their contribution to $b_{\rm res}$ when there is only one type of coupling at 
${\cal O}(\gamma)$ or lower and no parametric forcing $f_{2\Omega}$. In these cases 
there is no (leading order) dependence on the forcing phases. In the second section the 
same damped modes have been parametrically forced.  The factor $1/|\varrho|$ is then 
replaced by $P_{2\Omega} (\Phi)$ of (\ref{eq:P}), a strictly positive oscillatory function 
($|\varrho| > |\mu_if_{2\Omega}|$ for damped modes) with extrema at $\Phi = \pm 90^\circ$.  
Entries in the third section display two types of coupling, while the final two cases in the table 
have $f_{2\Omega}$ forcing as well.   Note that equivalent cases can be trivially generated 
from those in Table~\ref{tab:3ff} by switching $n$ and $p$ and relabeling, for example 
$(m,n,n-m)$, $\Omega=p$ is equivalent to $(m,n,m+n)$, $\Omega=n$.

Hamiltonian structure in the undamped problem has important consequences for the 
results in Table~\ref{tab:3ff}.  If one assumes that at $\gamma=0$ Eqs.~(\ref{eq:TW})  
derive from a Hamiltonian ${\cal H}$ through 
$d Z_j^\pm /d t = \mp i \partial {\cal H}/\partial \bar{Z}_j^\pm$ (see, \emph{e.g.}, \cite{Mile84}) 
then $q_1=q_5$, $q_2=q_7$, and $q_3=q_4=q_6$.   When we allow for simple rescalings: 
$(Z_1^\pm,Z_2^\pm) \rightarrow \eta (Z_1^\pm,Z_2^\pm)$, $(Z_3^\pm) \rightarrow \xi (Z_3^\pm)$ 
with $\eta,\xi \in {\mathbb R}$, these conditions relax to $q_1=r q_5$, $q_2=r q_7$, and 
$q_3=q_4=r q_6$ (for some $r >0$), implying
\begin{align}
\alpha_1 > 0, \quad \alpha_2 > 0, \quad \alpha_3 =0, \quad \alpha_4=0.
\label{eq:Ham}
\end{align}
Relations~(\ref{eq:Ham}) mean that for simple couplings (the first two sections 
of Table~\ref{tab:3ff}) the sign of $b_{\rm res}$ is determined, and thus one knows if the 
resonant triad has an enhancing or suppressing effect on patterns involving $\theta_{\rm res}$. 
The $\Omega=m$ and $\Omega = m \pm n$ modes are suppressing, the $\Omega=n$ mode is 
inconsequential, and the $\Omega=n-m$ mode is enhancing.

We now pursue the implications of Table~\ref{tab:3ff} by examining two cases relevant to 
recent experiments. We compare our theoretical predictions with coefficients calculated 
numerically from the Zhang-Vi\~nals Faraday wave equations \cite{ZhaVin97_1}, which 
describe weakly-damped fluids in deep containers. The calculation (see \cite{SilTopSke00} 
for details) gives us the cross-coupling coefficient $b=b_0 + b_{\rm res}$ at the SW bifurcation 
and is independent of the symmetry arguments used here.

\begin{figure}[ht]
\includegraphics[width=3.4in]{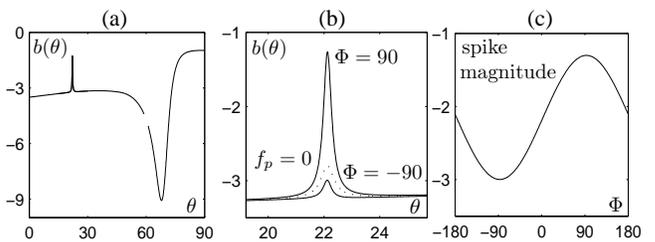}
\caption{Effect of $\Phi$ on the stabilizing $\Omega=n-m$ resonance that selects  
superlattice patterns \cite{KudPieGol98} with $(m,n,p)=(6,7,2)$ forcing. (a) $b(\theta)$ 
with the optimal phase $\Phi=90\degree$; the singular region heralding hexagons 
at $\theta=60\degree$ is removed.  The large dip near $\theta=67\degree$ degrees is due 
to the strongly suppressing $\Omega = m$ resonance in the first section of Table~\ref{tab:3ff}. 
(b) Close-up of $b(\theta)$ near $\theta=22\degree$ with $\Phi=90^\circ$ and 
$\Phi=-90^\circ$; the two-frequency result (dotted line) with $f_p=0$ is also shown. 
(c) Spike magnitude versus $\Phi$ (see Eq.~\ref{eq:P}).  For these calculations 
$|f_n|/|f_m|=0.75$, $|f_p|/|f_m|=0.1$.  The fluid parameters in the Zhang-Vi\~{n}als 
equations of \cite{ZhaVin97_1} are $\gamma = 0.08$, $G_0 = 1.5$.}
\label{fig:672}
\end{figure}

As our first example we consider the superlattice-I pattern observed with two-frequency 
$(m,n)=(6,7)$ forcing in the experiments of \cite{KudPieGol98}. The wave vectors making 
up the pattern lie on the vertices of two hexagons, one rotated by an angle 
$\theta_{\rm h} < 30^\circ$ with respect to the other (see Fig.~3 in \cite{SilTopSke00}). It 
was shown in \cite{SilTopSke00} that the experimentally observed angle of 
$\theta_{\rm h} \simeq 22^\circ$ is related to a resonant triad at $\theta \simeq  158^\circ$ 
($= 180^\circ - 22^\circ$) involving the $\Omega=n-m$ mode. This is the most interesting 
of the damped modes because it gives $b_{\rm res}>0$ and thus acts as a selection 
mechanism. Table \ref{tab:3ff} provides a means by which to enhance 
this effect, namely by parametrically forcing the damped mode via $(m,n,2n-2m)=(6,7,2)$ forcing. 
In Fig.~\ref{fig:672}a we show $b(\theta)$ for the phase $\Phi = 90\degree$ which optimizes the 
stabilizing effect at $\theta_h \simeq 22\degree$. In Fig.~\ref{fig:672}b we show that 
parametrically forcing the difference frequency mode with this phase quadruples the 
stabilizing impact with respect to the two-frequency case used in the experiments. Note 
that if the wrong phase ($\Phi=-90\degree$) is chosen, the effect of the resonance will 
actually be diminished compared to the two-frequency case. Fig.~\ref{fig:672}c shows the 
sinusoidal dependence of $b_{\rm res}$ on $\Phi$, in excellent agreement with our predictions.
%%%%%%%%%%%%%%%%%%%%%%%%%%%%%%%%%%%%%
%%%%%%%%%%%%%%%%%%%%%%%%%%%%%%%%%%%%%

As a second example, we consider recent experimental results on quasipatterns. It was reported 
in \cite{ArbFin02} that 8-fold quasipatterns,  which were distorted and difficult to observe with 
$(m,n)=(3,2)$ forcing, became dramatically cleaner and more robust with $(m,n,p)=(3,2,4)$ forcing. 
An explanation for this observation is provided by Table \ref{tab:3ff}. Specifically, we find that the 
$\Omega=1$ mode forms a resonant triad with the critical modes with associated angle 
$\theta_{\rm res} \approx 41\degree$. This is the angle present in the distorted (3,2)-forced 
quasipatterns in~\cite{ArbFin02}. Table \ref{tab:3ff} indicates that with $(m,n,p)=(3,2,4)$ 
forcing there is a positive $\alpha_2|f_4|^2$ contribution to $b_{\rm res}$ which has a stabilizing 
effect. In our numerical investigations, we find that the stabilizing spike in $b(\theta)$ becomes 
broader with increasing $\gamma$, and thus it is reasonable to expect that the stabilization could 
extend to the $45\degree$ angle associated with the perfect 8-fold quasipattern. Unfortunately, 
the experimental value of $\gamma$ appears to be too large for a more quantitative comparison 
with our theory.

The theoretical framework developed in this Letter should provide some welcome guidance for 
navigating the huge parameter space of multi-frequency forced Faraday waves. Table~\ref{tab:3ff} 
provides a comprehensive list of the important damped modes and their effect on pattern formation 
(additional possibilities with more than three frequencies will be considered elsewhere 
\cite{TopPorSil_pre}). In general, the influence of these damped modes, when parametrically 
forced,  depends on $T_\tau$-invariant combination(s) of forcing phases. Using the ``proper'' 
phase greatly enhances resonance effects while the ``wrong'' phase can actually reduce them. We 
hope that experimentalists will use Table~\ref{tab:3ff} as a tool to 
design forcing functions conducive to particular patterns.  For example, one might stabilize a 
quasipattern composed of 2N, ${\rm N}>3$, equally spaced critical modes by arranging for an 
$\Omega = n-m$ mode with wavenumber $k_3 = 2k_1\cos(\frac{{\rm N}-1}{\rm N} 90^\circ)$. 
In the inviscid limit an appropriate choice of $m$ and $n$ can be estimated from this geometric 
condition and the dispersion relation \cite{TopSil02}.   Adding a third $2n-2m$ frequency would 
then allow one to manipulate the effect by varying $\Phi$.  Finally, we emphasize that the scheme 
used to obtain Table~\ref{tab:3ff} is based on symmetry considerations and is therefore quite 
general, requiring only that the damping be small.

\begin{acknowledgements}
CMT was supported by NSF grants DMS-9983726 and DMS-9983320.  MS was supported 
by NASA grant NAG3-2364, NSF grant DMS-9972059, and by the NSF MRSEC Program 
under DMR-0213745.
\end{acknowledgements}
\bibliography{3ff}
\end{document}